\documentclass[twocolumn]{aastex631}

\shorttitle{Fine-structure constant with JWST}
\shortauthors{Jiang et al.}

\begin{document}

\title{Constraints on the variation of the fine-structure constant at $3<z<10$ with JWST emission-line galaxies}

\email{linhua.jiang@pku.edu.cn}

\author[0000-0003-4176-6486]{Linhua Jiang}
\affiliation{Department of Astronomy, School of Physics, Peking University, Beijing 100871, China}
\affiliation{Kavli Institute for Astronomy and Astrophysics, Peking University, Beijing 100871, China}

\author[0000-0003-0964-7188]{Shuqi Fu}
\affiliation{Department of Astronomy, School of Physics, Peking University, Beijing 100871, China}
\affiliation{Kavli Institute for Astronomy and Astrophysics, Peking University, Beijing 100871, China}

\author[0000-0002-7633-431X]{Feige Wang}
\affiliation{Steward Observatory, University of Arizona, 933 N Cherry Avenue, Tucson, AZ 85721, USA}

\author{Sarah E. I. Bosman}
\affiliation{Institute for Theoretical Physics, Heidelberg University, Philosophenweg 12, D–69120, Heidelberg, Germany}
 \affiliation{2 - Max-Planck-Institut f\"{u}r Astronomie, K\"{o}nigstuhl 17, 69117 Heidelberg, Germany}

\author[0000-0001-8467-6478]{Zheng Cai}
\affiliation{Department of Astronomy, Tsinghua University, Beijing 100084, China}

\author[0000-0003-1470-5901]{Hyunsung D. Jun}
\affiliation{Department of Physics, Northwestern College, 101 7th St SW, Orange City, IA 51041, USA}

\author[0000-0003-0230-6436]{Zhiwei Pan}
\affiliation{Department of Astronomy, School of Physics, Peking University, Beijing 100871, China}
\affiliation{Kavli Institute for Astronomy and Astrophysics, Peking University, Beijing 100871, China}

\author[0000-0002-4622-6617]{Fengwu Sun}
\affiliation{Steward Observatory, University of Arizona, 933 N Cherry Avenue, Tucson, AZ 85721, USA}
\affiliation{Center for Astrophysics $|$ Harvard \& Smithsonian, 60 Garden St., Cambridge, MA 02138, USA}

\author{Jinyi Yang}
\affiliation{Steward Observatory, University of Arizona, 933 N Cherry Avenue, Tucson, AZ 85721, USA}

\author{Huanian Zhang}
\affiliation{Department of Astronomy, Huazhong University of Science and Technology, Wuhan, Hubei 430074, China}

\begin{abstract}

We present constraints on the spacetime variation of the fine-structure constant $\alpha$ at redshifts $2.5\le z<9.5$ using JWST emission-line galaxies. The galaxy sample consists of 621 high-quality spectra with strong and narrow [\ion{O}{3}] $\lambda\lambda$4959,5007 doublet emission lines from 578 galaxies, including 232 spectra at $z>5$. The [\ion{O}{3}] doublet lines are arguably the best emission lines to probe the variation in $\alpha$. We divide our sample into six subsamples based on redshift and calculate the relative variation $\Delta\alpha/\alpha$ for the individual subsamples. The calculated $\Delta\alpha/\alpha$ values are consistent with zero within $1\sigma$ at all redshifts, suggesting no time variation in $\alpha$ above a level of $(1-2) \times10^{-4}$ ($1\sigma$) in the past 13.2 billion years. When the whole sample is combined, the constraint is improved to be $\Delta\alpha/\alpha = (0.2\pm0.7) \times10^{-4}$. We further test the spatial variation in $\alpha$ using four subsamples of galaxies in four different directions on the sky. The measured $\Delta\alpha/\alpha$ values are consistent with zero at a $1\sigma$ level of $\sim 2\times10^{-4}$. While the constraints in this work are not as stringent as those from lower-redshift quasar absorption lines in previous studies, this work uses an independent tracer and provides the first constraints on $\Delta\alpha/\alpha$ at the highest redshifts. With the growing number of emission-line galaxies from JWST, we expect to achieve stronger constraints in the future.

\end{abstract}


\section{Introduction} \label{sec:intro}

Fundamental physical constants do not vary in space and time in the framework of the Standard Model of particle physics. However, they are allowed or even required to vary in some modern theories beyond the Standard Model \citep{Martins2017}. Among these physical constants, the fine-structure constant $\alpha$ is of great interest. It is a dimensionless quantity that characterizes the strength of the electromagnetic interaction between elementary charged particles. In the past several decades, different methods have been developed to search for a possible variation in $\alpha$, including laboratory experiments and astrophysical observations \citep[e.g.,][]{Uzan2011}. In particular, astrophysical observations are able to constrain the $\alpha$ variation in the distant Universe. Previous studies have shown that any relative $\alpha$ variation $\Delta\alpha/\alpha$ or time variation {$\dot{\alpha} / \alpha$ ($\dot{\alpha} \equiv {\rm d}\alpha / {\rm d}t$), if it exists, must be extremely small \citep[e.g.,][]{Damour1996,Petrov2006,Rosenband2008,Murphy2022a}. 

The majority of the astrophysical observations in the past used quasar absorption lines. This is so far the primary method to probe the $\alpha$ variation at high redshift, and it relies on high-resolution, high signal-to-noise ratio (S/N) spectra of very bright quasars. The absorption lines used in early studies were usually fine-structure doublet lines such as \ion{C}{4}, \ion{N}{5}, \ion{Mg}{2}, and \ion{Si}{4} \citep[e.g.,][]{Potekhin1994,Cowie1995,Murphy2001_MN327_1237,Chand2005}. This so-called alkali-doublet (AD) method is clean and straightforward to trace the $\alpha$ variation. Later, the many-multiplet (MM) method was introduced to include more absorption lines \citep[e.g.,][]{Dzuba1999PhRvL_82_888,Webb1999}. This method takes advantage of different relativistic effects from different elements, and thus largely improves the detection sensitivity of $\Delta\alpha/\alpha$. On the other hand, it could suffer from a number of systematic uncertainties that have been demonstrated in the literature \citep[e.g.,][]{Murphy2001_MN_327_1223,Evans2014,Dumont2017,Lee2023,Webb2024}. Nevertheless, the MM method is currently the most widely-used technique and the strongest constraints on $\Delta\alpha/\alpha$ (null results) from this technique have reached an upper limit of several times $10^{-6}$ \citep[e.g.,][]{King2012,Molaro2013,Songaila2014,Wilczynska2015,Kotus2017,Milakovic2021,Murphy2022b,Lee2023,Webb2024}. 

Despite the fact that quasar absorption lines have been widely used in this field, the earliest astrophysical observations actually used quasar emission lines \citep[e.g.,][]{Savedoff1956,Bahcall1967,Bahcall2004,Albareti2015,Li2024}, particularly the [\ion{O}{3}] $\lambda\lambda$4959,5007 (hereafter [\ion{O}{3}]) doublet lines. Quasar emission lines are usually broad, and their wavelength measurement can be affected by a series of issues such as the contamination from the H$\beta$ and \ion{Fe}{2} emission near [\ion{O}{3}]. The great advantage of emission lines (compared to absorption lines) in astronomy is that it is much easier to observe with high-S/N spectra. Therefore, narrow and clean emission lines from galaxies potentially offer a great tool to search for a variable $\alpha$. Recently, \citet{Jiang2024} utilized $\sim110,000$ emission-line galaxies (ELGs) from the Dark Energy Spectroscopic Instrument \citep[DESI;][]{DESI2022_overview,DESI2023_EDR,DESI2023_Validation,Schlafly2023} to probe the spacetime variation of $\alpha$. The emission lines used in their work are the [\ion{O}{3}] doublet. This doublet is arguably the best choice among all emission lines, owing to its wide wavelength separation between the two doublet lines and its strong emission in many galaxies (see Section \ref{subsec:method}).

The \citet{Jiang2024} work based on the DESI data focused on a low-redshift range $z<1$, covering half of all cosmic time. Previous studies based on quasar absorption lines probed a redshift range of roughly $z \le 4$. At earlier times, a variable $\alpha$ is theoretically more possible \citep[e.g.,][]{Barrow2002,Alves2018}, but higher redshifts have been rarely explored \citep[e.g.,][]{Wilczynska2020}. Recently, JWST is collecting an unprecedented infrared dataset of galaxy and AGN spectra. An intriguing discovery is the existence of a high number density of strong [\ion{O}{3}] emitters at very high redshifts up to $z\sim10$ \citep[e.g.,][]{DEugenio2024,Heintz2024}. It allows us, for the first time, to explore the $\alpha$ variation at such high redshifts. 

In this work we constrain the $\alpha$ variation in the early epochs using a sample of 621 JWST spectra from 578 strong [\ion{O}{3}] emitters at $2.5 \le z<9.5$. The layout of the paper is as follows. In Section \ref{sec:data}, we introduce our methodology, spectroscopic data, and galaxy sample from JWST. In Section \ref{sec:waveCal}, we characterize the wavelength calibration of the JWST spectra. In Section \ref{sec:results}, we measure $\Delta\alpha/\alpha$ and present our results. We discuss the results and summarize the paper in Section \ref{sec:summary}. Throughout the paper, all magnitudes are expressed on the AB system. We use a $\Lambda$-dominated flat cosmology with $H_0=69\,{\rm km\,s}^{-1}\,{\rm Mpc}^{-1}$, $\Omega_{m}=0.3$, and $\Omega_{\Lambda}=0.7$.

\section{Methodology and spectral data} \label{sec:data}

In this section, we first introduce our method of calculating $\Delta\alpha/\alpha$ using the [\ion{O}{3}] emission lines. We then present our spectroscopic data collected from the JWST archive and the data reduction procedure. From these data, we address our target selection and build a large sample of [\ion{O}{3}] emission-line galaxies.

\subsection{Methodology} \label{subsec:method}

We calculate $\Delta\alpha/\alpha$ for individual spectra or galaxies using the [\ion{O}{3}] doublet lines based on the AD method. We use $\lambda_1$, $\lambda_2$, and $\bar{\lambda}$ to denote the two line wavelengths ($\lambda_1 < \lambda_2$) in vacuum and their average value. With a non-relativistic approximation for light elements, the wavelength separation (or difference) of the doublet lines $\lambda_{\rm D} = \lambda _2 - \lambda_1$ is directly related to $\alpha$ via $\lambda_{\rm D} / \bar{\lambda} \propto \alpha^2$ \citep{Uzan2003,Bahcall2004}. In order to calculate a small change of $\alpha$ (i.e., $\Delta\alpha$) between two redshifts $z_1$ and $z_2$, we take the derivative of the above relation and get $\Delta(\lambda_{\rm D} / \bar{\lambda}) \propto 2\alpha\,\Delta\alpha$. We then divide this new relation by the original relation, and obtain the following formula that has been frequently used in the literature,
\begin{equation}
\Delta\alpha / \alpha \approx 0.5 \times \left( \frac{\lambda_{\rm D}(z_2) / \bar{\lambda}(z_2)}{\lambda_{\rm D}(z_1) / \bar{\lambda}(z_1)} - 1 \right),
\end{equation}
where all wavelengths are in the rest frame. Equation 1 measures the relative $\alpha$ variation at two redshifts $z_1$ and $z_2$. One can see that any uniform shift of a spectrum caused by a cosmological redshift or a Doppler effect does not affect the $\Delta\alpha/\alpha$ measurement. 

In order to better illustrate how $\Delta\alpha/\alpha$ depends on $\lambda_{\rm D}$, we rewrite Equation 1 as follows. We assume $\bar{\lambda}(z_1) =  \bar{\lambda}(z_2) + \epsilon$, where $|\epsilon| \ll \bar{\lambda}$. We then replace $\bar{\lambda}(z_1)$ by $\bar{\lambda}(z_2)$ in Equation 1 and get
\begin{equation}
\Delta\alpha / \alpha \approx 0.5 \times \left( \frac{\lambda_{\rm D}(z_2) (1+\epsilon/\bar{\lambda}(z_2)) - \lambda_{\rm D}(z_1)} {\lambda_{\rm D}(z_1)} \right).
\end{equation}
We neglect $\epsilon/\bar{\lambda}(z_2)$ and obtain
\begin{equation}
\Delta\alpha / \alpha \approx 0.5\times \Delta(\lambda_{\rm D}) / \lambda_{\rm D}, 
\end{equation}
where $\Delta(\lambda_{\rm D})$ is the change of $\lambda_{\rm D}$, i.e., $\lambda_{\rm D}(z_2)-\lambda_{\rm D}(z_1)$ in this case. Equation 3 indicates that the wavelength separation $\lambda_{\rm D}$ is directly proportional to the sensitivity of the $\Delta\alpha/\alpha$ measurement. The wavelength separation of [\ion{O}{3}] in the rest frame is nearly 50 \AA, roughly one order of magnitude larger than most fine-structure doublet lines in the UV and optical. Owing to the wide wavelength separation and its strong emission in many galaxies, [\ion{O}{3}] is a powerful tool to probe the variation in $\alpha$. In the following analyses, we still use Equation 1 to calculate $\Delta\alpha/\alpha$.

We can also use the above equations to estimate the precision of the $\alpha$ measurement that we expect to achieve. Equation 3 shows that a $\lambda_{\rm D}$ change of $\Delta(\lambda_{\rm D})=0.01$ \AA\ for [\ion{O}{3}] implies $\Delta\alpha / \alpha \approx 10^{-4}$. In other words, a systematic or measurement uncertainty of 0.01 \AA\ from the relative wavelength calibration sets a detection limit of  $\Delta\alpha / \alpha \approx 10^{-4}$. As we will see in Section \ref{sec:results}, the accuracy of 0.01 \AA\ is roughly the best that we can achieve from our data. Because $\bar{\lambda}$ is about 100 times larger than $\lambda_{\rm D}$ (for [\ion{O}{3}]), $\Delta\alpha/\alpha$ is roughly 100 times less sensitive to the absolute wavelength calibration. This can be derived from Equation 1. In Equation 1, if we assume that $\lambda_{\rm D}$ does not change and we fix $\bar{\lambda}(z_1)$ as a reference value (see the last paragraph in this subsection), then an error propagation shows that an uncertainty of 1 \AA\ on $\bar{\lambda}(z_2)$ introduces an uncertainty of $10^{-4}$ to $\Delta\alpha/\alpha$. In reality, the absolute and relative wavelength calibrations are closely related.

From the above analysis or Equation 3, we can also obtain the required accuracy in the velocity space if we want to achieve an accuracy of $10^{-4}$ for $\Delta\alpha/\alpha$. Using $\Delta\lambda/\lambda = \Delta V/c$ ($c$ is speed of light) and assuming $\Delta\lambda=0.01$ \AA\ at $\lambda \approx 5000$ \AA, we obtain $\Delta V  \approx 600$ $\rm m\,s^{-1}$. This means that an accuracy of $\sim$600 $\rm m\,s^{-1}$ is needed from the relative wavelength calibration. 

Previous studies used the present-day wavelengths as the reference values. Following these studies, we assume $z_1=0$, then the two reference values are $\lambda_1(0) = 4960.295$ \AA\ and $\lambda_2(0) = 5008.240$ \AA, respectively\footnote{https://physics.nist.gov/PhysRefData/ASD/lines\_form.html}. These values are accurate enough for this work, and needs to be improved in the future. As we emphasized above, Equation 1 compares any two emission line systems at any two redshifts.

\subsection{Data and Data Reduction}

\begin{deluxetable*}{cccl}  
\tablewidth{0pt} 
\tablecaption{Spectral data compilation \label{tab:data}}
\tablehead{Instrument configuration & Data source & Number of spectra  & Program or PID}
\colnumbers
\startdata
NIRSPEC/MOS (G235M/F170LP) & JADES  & 268 &  1180, 1210, 1286 \\
NIRSPEC/MOS (G235M/F170LP) & CEERS & 47  &  1345 \\
NIRSPEC/MOS (G235M/F170LP) & DJA       & 52  &  1181, 1210, 1345, 2593, 2736, 4446 \\
NIRSPEC/MOS (G235H/F170LP) & DJA       & 19   &  1324 \\
NIRSPEC/MOS (G395M/F290LP) & JADES  & 140 &  1180, 1210, 1286, 3215 \\
NIRSPEC/MOS (G395M/F290LP) & CEERS & 38   &  1345 \\
NIRSPEC/MOS (G395M/F290LP) & DJA       & 29   &  1210, 2674, 2736 \\
NIRSPEC/MOS (G395H/F290LP) & JADES  & 18   &  1210 \\
NIRSPEC/MOS (G395H/F290LP) & DJA       & 10   &  1210, 1324\\
\enddata
\tablecomments{JADES spectra dominate our sample. Column 3 indicates the number of spectra that are actually used in this work.}
\end{deluxetable*} 

\begin{figure}[t]   
\includegraphics[width=0.45\textwidth]{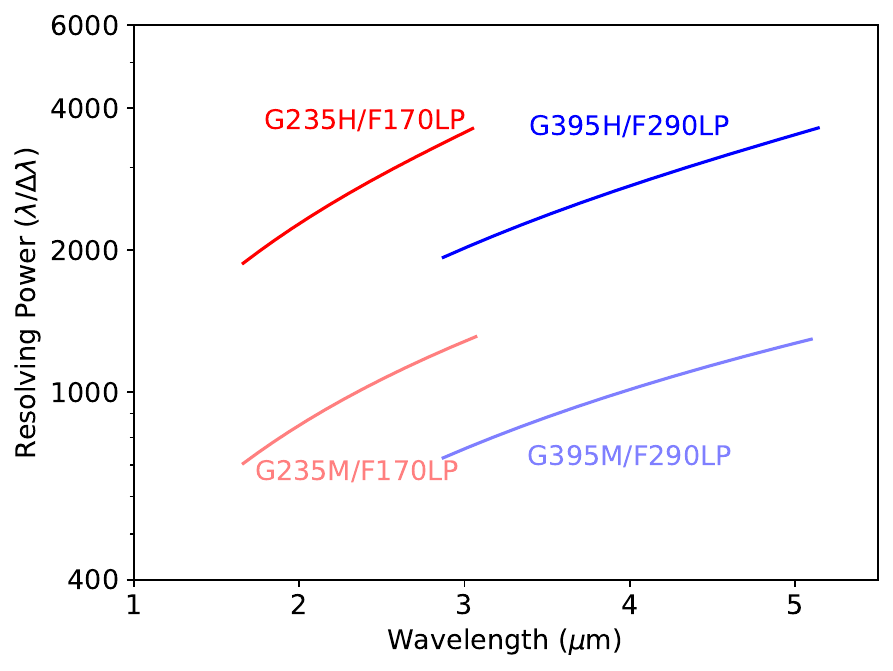}
\caption{JWST NIRSpec configurations (combinations of dispersers/grating and filters) used for this work. The color coded curves represent four configurations of the NIRSpec MOS mode. \label{fig:resolution} }
\end{figure}

The spectral data used in this work were obtained from multiple JWST programs (Table~\ref{tab:data}). They were observed by the NIRSpec multi-object spectroscopy (MOS). We have also considered data observed by the NIRCam wide field slitless spectroscopy (WFSS). After we analyze their wavelength calibration (see details in Section \ref{sec:waveCal}), we find that the wavelength measurement of the WFSS spectra does not satisfy our requirement. In this subsection, we introduce both NIRSpec MOS and NIRCam WFSS data, but we only use NIRSpec spectra to measure variation in $\alpha$. 

NIRSpec is one of the four instruments on board JWST. It is a versatile spectrograph with different observing modes. In particular, the MOS mode using micro shutter array (MSA) can observe up to 200 objects simultaneously \citep{Ferruit2022}, and is thus very powerful to observe high-redshift galaxies. The spectra used in this work were made with NIRSpec MOS. The MOS mode provides 8 configurations (combinations of dispersers/gratings and filters) that cover a wide wavelength range of $0.7-5.3$ \micron. Different configurations have different resolving powers ($R\sim700-4000$). Figure \ref{fig:resolution} illustrates four NIRSpec configurations used in this work. The figure also shows that there are two sets of gratings, including medium-resolution gratings with $R\sim1000$ denoted by `M' and high-resolution gratings with $R\sim2700$ denoted by `H'. In addition, the resolving power for each individual configuration changes significantly with wavelength. The pixel scales for the two medium-resolution gratings G235M and G395M in the figure are 10.68 and 17.95 \AA, respectively. This means that the medium-resolution spectra are just critically sampled. For example, for the G395M grating at 4 $\micron$, $R\sim1000$ indicates a resolution of $\Delta \lambda \sim 40$ \AA, which is roughly twice the pixel scale. The vast majority of our spectra were observed with these two medium-resolution gratings. For the two high-resolution gratings G235H and G395H, their pixel scales are 3.77 and 6.65 \AA, respectively. NIRSpec also offers prism observations. We did not use the prism data because of their very low spectral resolution. 

Two of the largest JWST Cycle 1 NIRSpec programs are the JWST Advanced Deep Extragalactic Survey \citep[JADES;][]{Eisenstein2023} and the Cosmic Evolution Early Release Science Survey \citep[CEERS;][]{Finkelstein2023}. The JADES program observes two deep pointings and 14 medium-deep pointings in the two GOODS fields, GOODS-S and GOODS-N. We included data from proposal IDs (PIDs) 1180, 1181, 1210, 1286, and 3215. As shown in Table~\ref{tab:data}, the JADES spectra dominate our sample. The CEERS program ( PID: 1345) observes six pointings in the Extended Groth Strip (EGS) CANDELS field. The details of the data reduction are presented in \citet{Arrabal_Haro2023} and \citet{Bunker2023}. The reduction procedure includes  the subtraction of dark current and bias, the removal of $1/ f$ noise and snowballs, flat-fielding, background subtraction, photometry, wavelength calculation, and slit-loss correction, etc. The individual 2D spectra of each target were then rectified and combined to generate the final 2D spectrum. In addition to the two programs, we collected more high-level NIRSpec spectra from the DAWN JWST Archive (DJA) \citep{Heintz2024}. The data were reduced with MsaExp \citep{Brammer2022}, and the following five programs were included: 1324, 1345, 2674, 2736, and 4446 (see also Table~\ref{tab:data}). The wavelength scales per pixel in the final data products are the same as the native scales. All above data were obtained before May 2024.

NIRCam is the major imaging (for $<5$ \micron) instrument on JWST. It also provides a powerful capability for slitless grism spectroscopy, i.e., WFSS  \citep{Greene2017}. NIRCam WFSS covers a wavelength range of $2.4-5.0$ \micron\ when used with different filters. Its resolving powers is roughly $R\sim1500$, and its pixel scale is about 10 \AA\ per pixel. We used WFSS spectra from two JWST programs, ``Medium-band Astrophysics with the Grism of NIRCam in Frontier Fields" (MAGNIF; PID: 2883) and ``A SPectroscopic Survey of Biased Halos in the Reionization Era" (ASPIRE; PID: 2078). The MAGNIF program makes WFSS observations in the Frontier Fields using the F360M filter in the long wavelength channel and the F182M filter in the short wavelength channel, simultaneously. We included the Abell 2744 (A2744) cluster field here. The ASPIRE program observes 25 high-redshift ($z>6.5$) quasar fields, and uses the F356W filter for WFSS observations and the F200W filter for direct imaging observations. The details can be found in \citet{Wang2023}. 

The WFSS data were reduced as follows. They were first reduced to the level of Stage-1 using the standard JWST calibration pipeline\footnote{https://jwst-docs.stsci.edu/jwst-science-calibration-pipeline}. We applied a flat-field correction using flat-field images taken with the same filter and module, and subtracted the $1/f$ noise along rows for grism-C exposures. We then performed a 2D sky-background subtraction using sigma-clipped median images. The WCS of each image was calibrated with the Gaia DR3 catalog \citep{Gaia2023} by matching stars detected in the short wavelength images. We extracted 2D spectra from individual exposures and stacked the 2D spectra after registering them to a common wavelength and spatial grid. Finally, 1D spectra were extracted from the stacked 2D spectra \citep{Sun2023,Wang2023}. The wavelength scales per pixel in the final data products are the same as the native scales.

\subsection{Galaxy Sample}

From the above JWST data, we searched the spectra for [\ion{O}{3}] doublet emission lines with high S/Ns and built a sample of bright [\ion{O}{3}] emitters. The line search was based on the detection significance of line emission, not the observed line flux density or absolute line luminosity, because different programs have different exposure times. We used a line-searching algorithm based on the method by \citet{Wang2023}. The procedure was straightforward. We required that each [\ion{O}{3}] $\lambda$4959 line should be detected above a $7\sigma$ level. Note that the [\ion{O}{3}] $\lambda$5007 line is about three times stronger than the [\ion{O}{3}] $\lambda$4959 line. Fainter lines do not help us improve our constraint on $\Delta\alpha/\alpha$. Since our targets are bright doublet emission lines, contaminant lines can be easily identified and rejected. 

Some galaxies have more than one spectrum for two reasons. One reason is that the wavelength coverages of the G235 and G395 gratings slightly overlap. The other reason is that some galaxies were observed in both medium- and high-resolution modes. Our analyses are based on spectra (not galaxies), i.e., two spectra of one galaxy are treated independently. This is consistent with previous studies based on quasar absorption lines, in which a small number of very bright quasars were observed multiple times by different telescopes and instruments in multiple years, and each observation was considered as an independent measurement. 

We visually inspected the 1D spectra of the [\ion{O}{3}] emitter candidates and removed lines with apparently asymmetric  features or candidates whose two doublet lines have apparently different shapes due to cosmic rays, contamination from nearby objects, or unknown reasons. Our final sample consists of 621 spectra from 578 strong [\ion{O}{3}]-emitting galaxies, including 232 spectra at $z>5$. The sample is shown in Table \ref{tab:sample} (the whole table is available on line). Among these galaxies, 42 galaxies have two spectra and one galaxy has three spectra. In this work we focus on high redshift, since the low-redshift regime has been well explored by the quasar absorption line method. Figure \ref{fig:redshift} shows the redshift distribution of our spectra in a redshift range of $2.5\le z<9.5$. The redshifts are from the public catalogs. The highest redshift in our sample is 9.43, when the Universe was only 0.5 Gyr old. As we mentioned before, these spectra are all from NIRSpec MOS observations. We emphasize that `a strong line' means a line with a high S/N, not necessarily a high flux density, as we explained earlier. 

\begin{figure}[t]   
\includegraphics[width=0.45\textwidth]{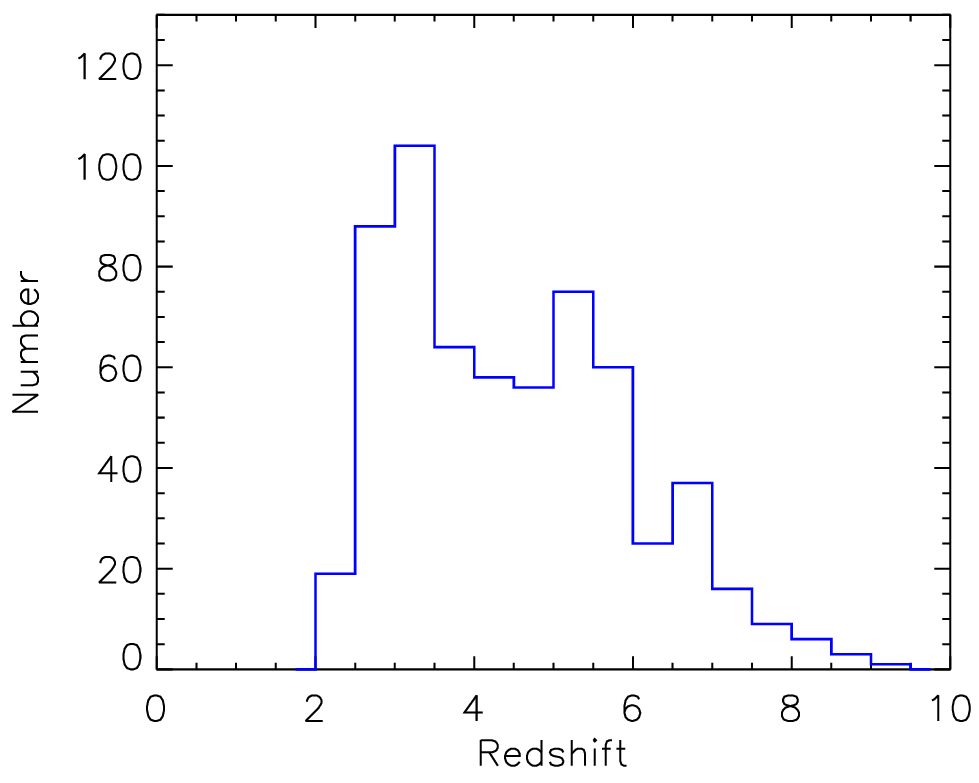}
\caption{Redshift distribution of 621 spectra (578 galaxies) with high S/Ns in our sample. The sample covers a redshift range of $2.5 \le z<9.5$, with 232 spectra at $z>5$. \label{fig:redshift} }
\end{figure}

\begin{deluxetable*}{ccccccccc}  
\tablewidth{0pt} 
\tablecaption{Galaxy/spectral sample used in this work \label{tab:sample}}
\tablehead{No. & R.A. (J2000.0) & Decl. (J2000.0) & Redshift & Data source & Grating & FWHM (\AA) & Flux ratio & $\Delta\alpha/\alpha$ ($\times 10^{-3}$) }
\colnumbers
\startdata
  1 & 03:32:36.024 & --27:49:10.96 &   3.4680 & JADES & G235M &  4.39 &  2.69 &  0.155$\pm$1.432 \\
  2 & 03:32:39.014 & --27:48:59.80 &   2.8070 & JADES & G235M &  5.76 &  3.02 & --1.036$\pm$0.857 \\
  3 & 03:32:40.490 & --27:48:54.58 &   4.0230 & JADES & G235M &  5.52 &  2.69 & --1.540$\pm$0.944 \\
  4 & 03:32:37.315 & --27:48:13.97 &   4.0440 & JADES & G235M &  4.37 &  3.01 & --0.792$\pm$2.127 \\
  5 & 03:32:27.199 & --27:48:10.76 &   4.1490 & JADES & G235M &  4.14 &  2.86 &  0.808$\pm$0.814 \\
  6 & 03:32:31.133 & --27:48:09.86 &   3.5780 & JADES & G235M &  5.26 &  2.16 &  0.272$\pm$2.473 \\
  7 & 03:32:36.007 & --27:48:09.00 &   4.2290 & JADES & G235M &  4.28 &  2.62 &  0.855$\pm$1.120 \\
  8 & 03:32:39.043 & --27:48:08.53 &   4.8060 & JADES & G235M &  4.21 &  2.68 &  1.516$\pm$1.197 \\
  9 & 03:32:40.183 & --27:46:31.73 &   4.1340 & JADES & G235M &  4.25 &  2.41 &  2.596$\pm$2.850 \\
 10 & 03:32:40.123 & --27:46:28.63 &   4.7750 & JADES & G235M &  4.18 &  2.91 & --1.493$\pm$0.894 \\
 \enddata
\tablecomments{Column 7 shows the measured FWHM of the [\ion{O}{3}] $\lambda$5007 line at rest frame , without correction for the instrumental broadening. Column 8 shows the flux ratio of the doublet lines. The whole table is available on line.}
\end{deluxetable*}

\section{Wavelength Calibration} \label{sec:waveCal}

Wavelength calibration is a key step to achieve the high accuracy of wavelength measurements, so previous studies paid close attention to it. For ground-based observations, wavelength calibration data (usually arc lamp spectra) are often taken immediately before or after science exposures. However, it is unrealistic for JWST to perform such a calibration, since there are no arc lamps on board JWST. In addition, JWST rarely takes calibration data to save time, because JWST major science goals do not desire a very accurate wavelength calibration. Therefore, we use indirect methods to characterize the wavelength calibration. Specifically, we use emission lines from objects observed by the instruments.

\subsection{NIRSpec MOS Wavelength Calibration}

In this and the next subsections, we estimate the systematic uncertainties of the wavelength calibration for the NIRSpec MOS data. We do this for each individual configuration and each individual data product (JADES, CEERS, and DJA; see Table~\ref{tab:calib}). As mentioned above, it is very challenging for JWST to achieve an accurate wavelength measurement. It is also unrealistic to characterize systematics for individual objects or spectra, not only because of the lack of immediate wavelength calibration data, but also due to other complex issues. Here are two major known issues that mainly affect the absolute wavelength calibration. The first one is from the instrument design. As mentioned by \citet{Alves_de_Oliveira2018}, each time when the instrument comes back to a given disperser or to the mirror, it stops at a slightly different position, which may slightly change the wavelength zero point. The second one originates from the nature of high-redshift objects. Many high-redshift galaxies are very compact and can be smaller than the slit width (which is fixed). This slit effect also exists in ground-based observations, but is more severe in JWST data because of their exceptional PSF quality. It is nearly impossible to correct this slit effect for individual objects, because our study relies on emission lines, and the emission line region in a galaxy is usually unknown from its broad-band images that are used to put the galaxy in a slit. Therefore, we can only statistically estimate the systematics of the wavelength measurements.

The absolute and relative wavelength calibrations are closely related, and an uncertainty in the absolute calibration propagates to the relative calibration. This can be illustrated by a simple case in which there is a pair of emission lines at rest-frame 5000.0 and 5100.0 \AA, respectively. If this system is at $z=0.0001$, then the observed wavelengths would be 5000.50 and 5100.51 \AA. If this system is at $z=0$, but suffers from a global wavelength drift of 0.5 \AA, then the observed wavelengths would be 5000.50 and 5100.50 \AA. In this case, the drift of 0.5 \AA\ causes a bias of 0.01 \AA\ in the relative wavelength measurement. This case is similar to our JWST data. There are often global velocity drifts in JWST spectra \citep[e.g.,][]{DEugenio2024} due to reasons including the two issues given in the above paragraph. They cause zero-point drifts in the absolute wavelength calibration, so we first briefly assess the absolute calibration below. 

From the JWST spectra in Section 2, we select a sample of galaxies with strong emission lines and cross match these galaxies with the VANDELS catalog \citep{Garilli2021}. The VANDELS survey is an optical spectroscopic survey of $1<z<6.5$ galaxies and the majority of these galaxies are bright with $i<25$ mag. We only use their spectra with the best quality (quality flags 3 or 4) in the VANDELS catalog. Based on \citet{Garilli2021}, the wavelength calibration uncertainty of the VANDELS spectra is below 0.4 \AA. Because the JWST targets are generally much fainter, there are only a small number of galaxies in common, and they are mostly at redshifts $2.4<z<4.8$. 
For each galaxy in the matched sample, we compare its redshifts from VANDELS and from the JWST infrared spectra. The average redshift difference is about 0.002 (or $\Delta z / (1+z) \sim 0.0005$), suggesting that the two measurements are generally consistent. \citet{DEugenio2024} has also investigated the wavelength calibration for the JADES spectra. They found an average velocity drift of $\sim$30 $\rm km\, s^{-1}$ in the wavelength solution, corresponding to 0.5 \AA\ at rest-frame 5000 \AA. After they corrected this bias, the residuals are smaller. Therefore, we conclude that the absolute wavelength calibration of the JWST programs that we used is robust.

\subsection{Relative Wavelength Calibration}

We then characterize the relative wavelength calibration of NIRSpec MOS. From the data described in Section \ref{sec:data}, we collect all spectra that have two or more strong emission lines ($\rm S/N > 10$). The Ly$\alpha$ line is usually asymmetric due to the IGM absorption and thus not considered. Doublet lines such as [\ion{C}{4}] $\lambda\lambda$1548,1550 are barely resolved in the spectra and not considered as well. In addition, the [\ion{O}{3}] doublet lines are not used. In the final collection, most spectra have 2-4 strong emission lines. For a given spectrum, its emission lines form one or more pairs. For each line pair, we compare its theoretical and observed wavelength separations as follows. All wavelengths in this subsection are expressed in the observed frame. We use the line at the shorter wavelength to calculate a redshift (see the next section for how we measure a line wavelength), and use this redshift to predict the wavelength of the other line. The difference between the predicted and observed wavelengths, denoted as $\delta\lambda$ here, is caused by the combination of the measurement and systematic uncertainties. We hope to use the large spectral sample to suppress the measurement uncertainties. Because different line pairs have very different wavelength separations ($\lambda_{\rm diff1}$), we cannot directly use their $\delta\lambda$ values. Therefore, we scale $\delta\lambda$ using $\delta\lambda / \lambda_{\rm diff1} \times \lambda_{\rm diff2}$, where $\lambda_{\rm diff2}$ is the line wavelength separation of a hypothetical [\ion{O}{3}] doublet that is put at the mean wavelength of the line pair. The underline assumption is that $\delta\lambda$ changes linearly with the wavelength separation. We are not able to determine higher-order distortions. Under this assumption, the scaled $\delta\lambda$ reflects the uncertainty of the wavelength calibration for [\ion{O}{3}]. 

\begin{figure}[t]   
\includegraphics[width=0.45\textwidth]{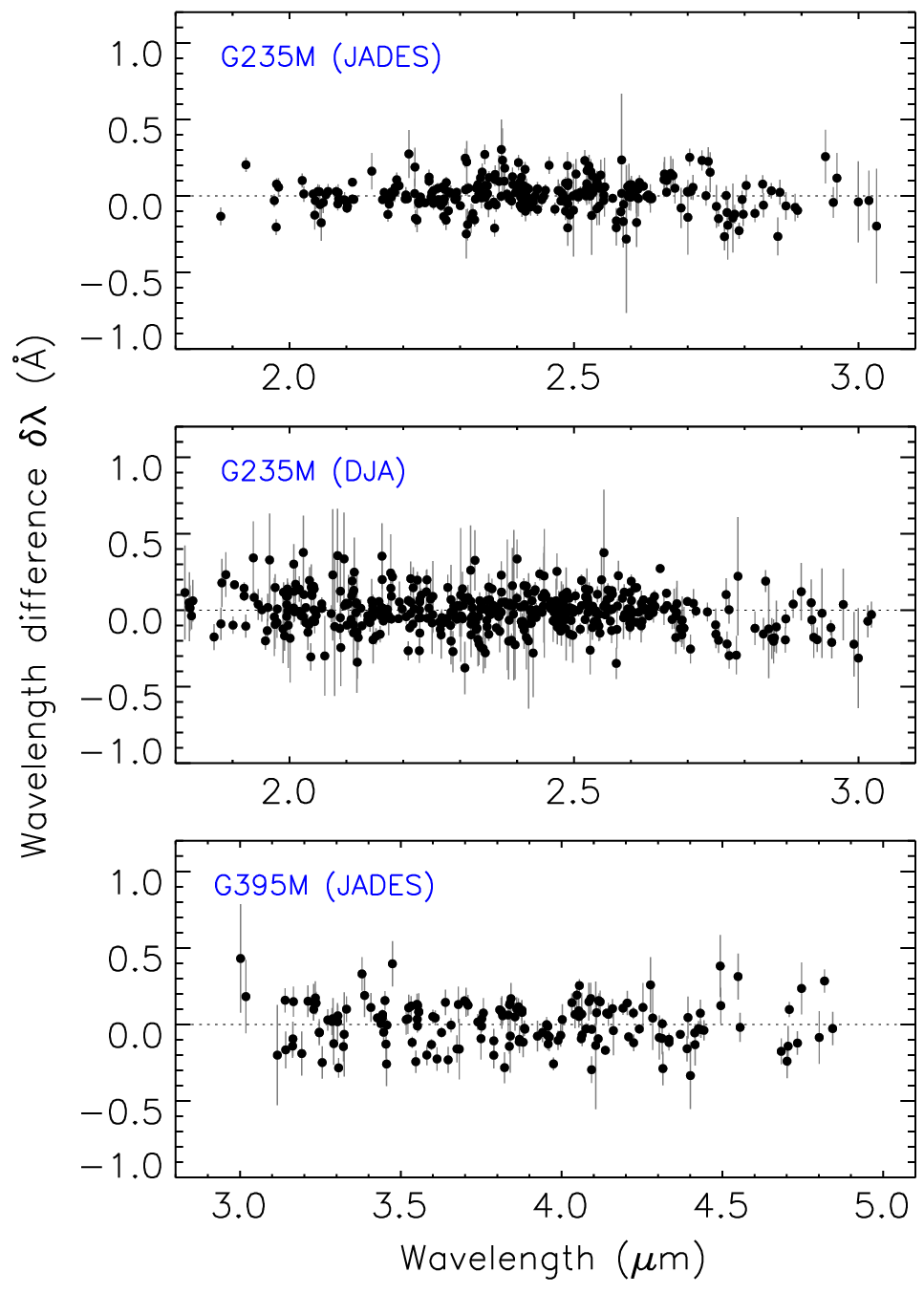}
\caption{Wavelength differences ($\delta\lambda$) of emission line pairs used for wavelength calibration. The wavelengths here are in the observed frame. The filled circles represent individual line pairs and their error bars indicate $1\sigma$ uncertainties from the wavelength measurement. The wavelength difference of a line pair is defined by the difference of the observed and theoretical wavelength separations of the two lines, scaled by the wavelength separation of a hypothetical [\ion{O}{3}] doublet at the same redshift (see Section 3.2 for details). \label{fig:wCalib1} }
\end{figure}

\begin{figure}[t]   
\includegraphics[width=0.45\textwidth]{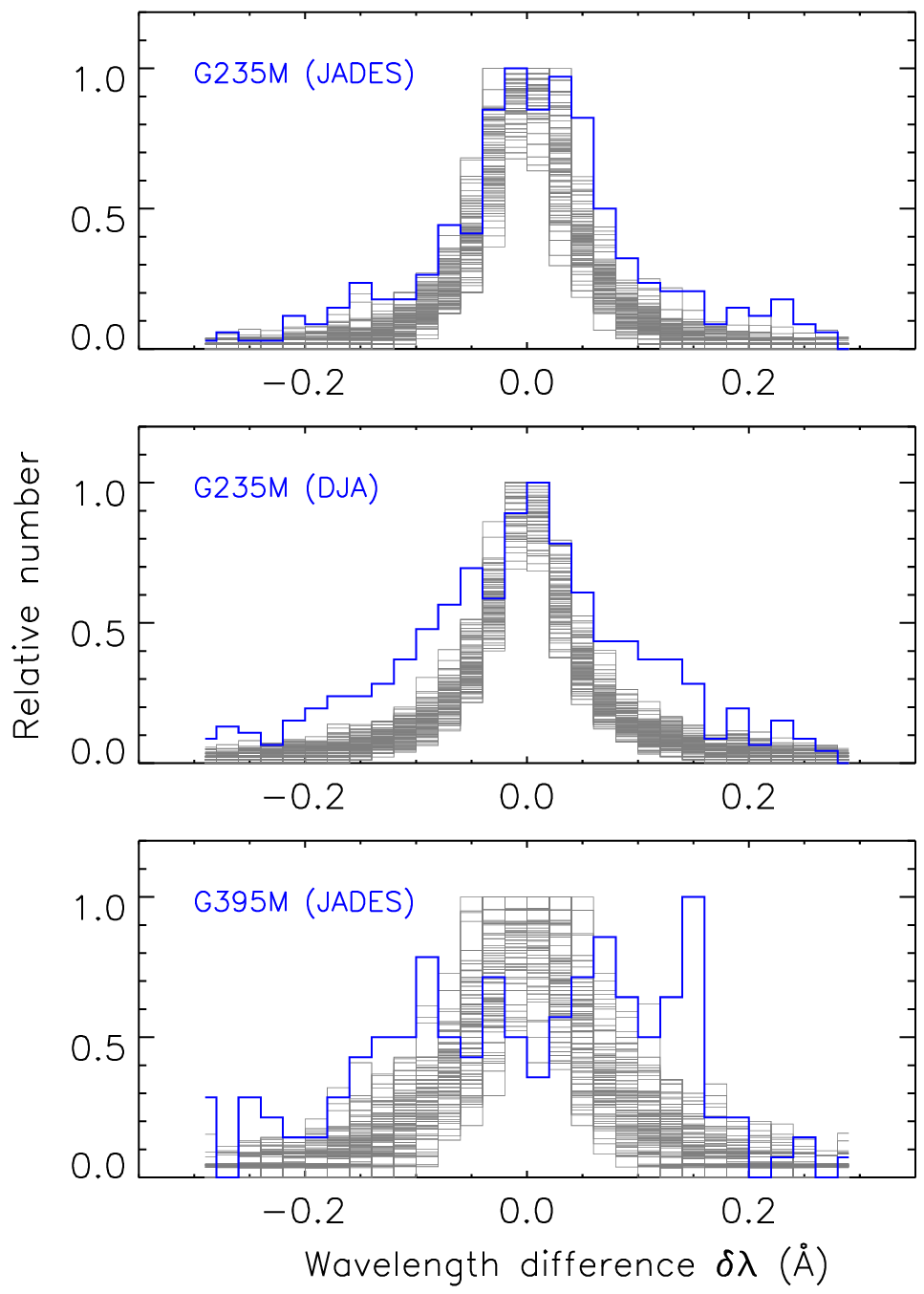}
\caption{Distributions of the wavelength differences ($\delta\lambda$) from the observed and simulated data addressed in Section 3.2. The  wavelengths here are in the observed frame. In each panel, the blue histogram represents the distribution of the data points from the corresponding panel in Figure \ref{fig:wCalib1}. The 100 gray histograms represent the distributions of the 100 simulated datasets that contain the measurement errors only. All histograms have been scaled so that their peak values are one. Compared with the distributions of the simulated data, if the distribution of the observed distribution is wider, the systematic errors are more important. \label{fig:wCalib2} }
\end{figure}

Figure \ref{fig:wCalib1} illustrates the scaled $\delta\lambda$ for two representative data products/sources (JADES and DJA) and two configurations (or gratings, including G235M and G395M). The error bars in the figure are $1\sigma$ uncertainties from the wavelength measurement. The blue histograms in figure \ref{fig:wCalib2} show the distributions of $\delta\lambda$ corresponding to the three panels in Figure \ref{fig:wCalib1}. Table \ref{tab:calib} summarizes the statistical results of the scaled $\delta\lambda$ for all combinations of gratings and data sources used in this work. The third and fourth columns of the table list the direct median and standard deviation of the $\delta\lambda$ distribution (as seen in Figure \ref{fig:wCalib2}). The last column is the weighted mean and its $1\sigma$ error. A weighted mean is defined as $\Sigma(x_i / \sigma_i^2) / \Sigma(1/\sigma_i^2)$, where $x_i$ and $\sigma_i$ are individual measurements and their errors. We can see that either mean or median values are very small. As mentioned earlier, the uncertainties from the absolute wavelength calibration have been propagated into the above calculation. We use the values in this table to correct our results in Section \ref{sec:results} as follows. 

We assume that the systematics consist of two parts, a constant wavelength shift and a scatter on the wavelength measurement, and we discuss them separately. The major effect of systematics is usually a constant shift. The median values in Table \ref{tab:calib} are used to correct for the constant shift, assuming that these values are dominated by systematics. On the other hand, the comparison between the median and mean values suggests that the measurement errors are not completely negligible for all cases. Therefore, we perform such a correction only if the absolute value of a median is greater than 0.02 \AA. This is also to avoid potential overcorrections. Specifically, we perform a constant correction for G395M+JADES and G395M+CEERS only. 

We then correct the second systematic effect that may have produced a scatter on the wavelength measurement. We use Monte Carlo simulations to check whether the scatters of the $\delta\lambda$ distributions originate from the measurement errors only. For example, for each data point in a panel of Figure \ref{fig:wCalib1}, we randomly generate 100 values whose mean value is zero and standard deviation is the $1\sigma$ error of this data point. The distributions of these 100 simulated datasets are shown as the 100 gray histograms in Figure \ref{fig:wCalib2}. These histograms largely overlap, but their overall profiles are obvious. 

In the top panel of Figure \ref{fig:wCalib2}, the observed $\delta\lambda$ distribution is roughly consistent with the simulated distributions, suggesting that the scatter of the $\delta\lambda$ distribution is dominated by the measurement errors. In the middle panel, the observed distribution is apparently wider than the simulated distributions. In other words, the measurement errors alone cannot explain the scatter of the distribution. This suggests the existence of significant systematic uncertainties. The difference between the two panels also reflects the heterogeneity of the data products. Our spectra were from three difference sources and were processed by different data reduction pipelines and methodologies. For instance, JADES and CEERS mostly rely on the standard JWST pipeline, whereas DJA largely uses a custom software. In addition, JADES optimizes the reduction steps for background subtraction, spectral rectification, 1D optimal extraction, etc. \citep{DEugenio2024}. The usage of different data products, on one hand, allows us to crosscheck our results. On the other hand, it may introduce potential systematics. This is also one reason that we characterize the wavelength calibration for each individual configuration and each individual data product, so that we can handle potential systematics individually (see the following analyses). The difference between the two panels can be explained by the fact that JADES has applied a correction for the slit effect. As mentioned earlier, the slit effect increases the uncertainties of the wavelength zero point values. This effect has been largely reduced by JADES.

In the bottom panel of Figure \ref{fig:wCalib2}, the observed $\delta\lambda$ distribution marginally agrees with the simulated distributions. However, they are all wider (scatters are larger) than those shown in the above two panels. There are a few reasons. First of all, the G395M grating covers longer wavelengths and the wavelength calibration becomes more difficult. Second, this grating corresponds to higher-redshift galaxies. Higher-redshift galaxies tend to be fainter than lower-redshift galaxies, so their lines are weaker (lower S/Ns) in the spectra. Another reason is that the $\delta\lambda$ values have been scaled to match the observed wavelength separation of [\ion{O}{3}], and thus the scale factors ($\lambda_{\rm diff2}$) are larger at longer wavelengths.

\begin{deluxetable*}{ccccc}  
\tablewidth{0pt} 
\tablecaption{Results of wavelength calibration \label{tab:calib}}
\tablehead{Instrument configuration & Data source & Median ($\delta\lambda$)  & Standard deviation ($\delta\lambda$) & Weighted mean ($\delta\lambda$)}
\colnumbers
\startdata
G235M/F170LP & JADES  &  0.009     &  0.102  &  $0.002\pm0.002$ \\
G235M/F170LP & CEERS &  0.001     &  0.129  &  $-0.011\pm0.004$ \\
G235M/F170LP & DJA       & $-0.001$ &  0.126  &  $0.007\pm0.001$ \\
G235H/F170LP & DJA       &  0.004     &  0.080  &   $0.003\pm0.002$ \\
G395M/F290LP & JADES  &  0.020     &  0.144  &   $0.031\pm0.004$ \\
G395M/F290LP & CEERS & $-0.033$ & 0.191   &   $0.010\pm0.011$ \\
G395M/F290LP & DJA       &  0.002     & 0.262   &   $0.022\pm0.004$ \\
G395H/F290LP & JADES  &  0.004      &  0.133  &   $0.004\pm0.006$ \\
G395H/F290LP & DJA       &  0.0      &  0.162  &   $0.014\pm0.003$  \\
\enddata
\tablecomments{The units for Columns 3, 4, and 5 are \AA.}
\end{deluxetable*} 

Based on the above analyses, we perform a correction for the systematics using the comparison between the observed and simulated $\delta\lambda$ distributions. We make this correction only if the observed distribution is apparently wider to avoid potential overcorrections. This does not mean that this effect, if existing, is very small for all other configurations. This merely means that in other cases the measurement errors are large and this systematic effect is relatively less important. In order to apply the correction, we assume that this effect is stochastic and Gaussian distributed. Usually systematic errors are neither stochastic nor Gaussian. But this is an acceptable approximation for our data. As mentioned in Section 3.1, the major known systematic uncertainties originate from the uncertainties of the wavelength zero points. Due to the instrumental effect and the nature of our objects, we are not able to determine precise zero points for individual objects. But it is reasonable to assume that the measured zero points are randomly distributed around the true values, so this systematic error is added in quadrature to the measurement error. Nevertheless, this correction is applied to only 19 objects in our sample, and the overall correction is small.

From the above analyses, the NIRSpec wavelength calibration is generally accurate and precise. The systematic errors that we have identified are small and their impact on our final result is limited compared with the measurement errors (see the next section). JWST is operating in space and its observing conditions are extremely stable, so the spectral dispersion should be very stable, which would result in a stable relative wavelength calibration. Therefore, we suspect that the above systematics are dominated by the absolute wavelength calibration (particularly the uncertainties of the wavelength zero points). High S/N calibration data are needed in the future to fully characterize the NIRSpec wavelength calibration.

\subsection{NIRCam WFSS Wavelength Calibration}

We carry out the same analyses for the NIRCam WFSS wavelength calibration. Specifically, we use known emission line galaxies from ground-based observations to estimate the absolute calibration, and use emission line pairs identified from the WFSS spectra to estimate the relative calibration. We find that the calibration accuracy is more than five times worse than the NIRSpec data. This is consistent with previous findings in the literature \citep[e.g.,][]{Sun2023,Torralba-Torregrosa2024}. The main reason is well known, i.e., it is impossible to accurately determine the centroids of line-emitting regions from the current imaging data for slitless spectra. Given that the number of the WFSS spectra is less than 10\% of the total spectra, we do not use the WFSS spectra in the following calculations of $\Delta\alpha/\alpha$.

\section{Results} \label{sec:results}

In this section we measure $\Delta\alpha/\alpha$ and constrain the spacetime variation of $\alpha$ at high redshift. We basically follow the procedure in our previous work \citep{Jiang2024}. Compared to the DESI data in \citet{Jiang2024}, the galaxy sample in this work is much smaller, and the spectral resolutions and S/Ns of the lines are lower. Therefore, we expect that the final $\Delta\alpha/\alpha$ uncertainties would be much larger.

\subsection{Measurement of $\Delta\alpha/\alpha$} \label{subsec:calc_alpha}

\begin{figure}[t]   
\includegraphics[width=0.45\textwidth]{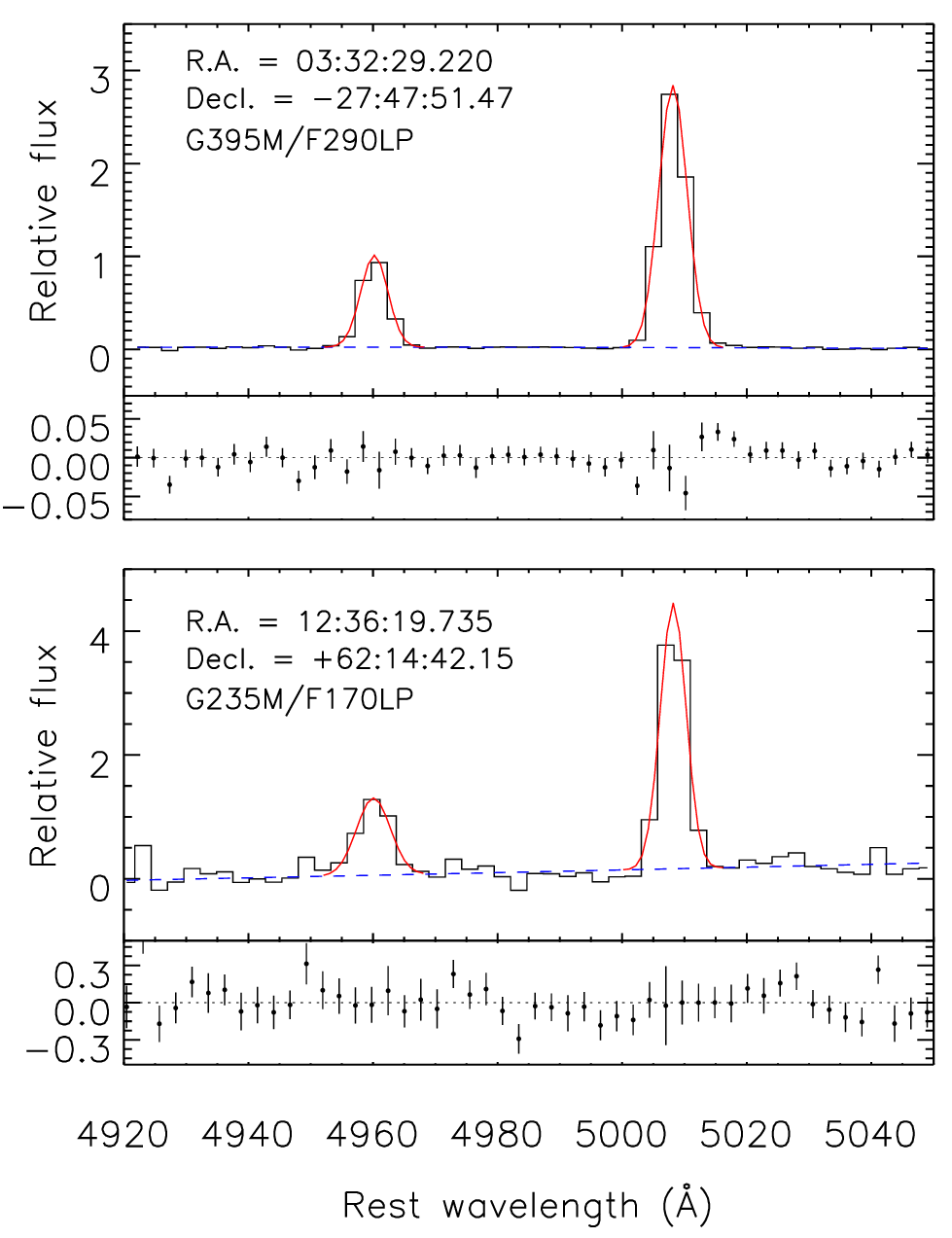}
\caption{Examples of a single Gaussian fit to [\ion{O}{3}] in two spectra. In each panel, the black spectrum has been shifted to the rest frame. The dashed line indicates the zero value. The best-fit continuum is in blue. Note that the majority of the strong line emitters in our sample have very weak continuum emission that is barely seen in the NIRSpec spectra. The two best-fit Gaussian profiles (plus underline continuum emission) are in red. The lower sub-panel shows the fitting residual, and the error bars indicate the $1\sigma$ errors of the data points from the spectrum. \label{fig:1gfit} }
\end{figure}

We measure the wavelengths of the emission lines and calculate $\Delta\alpha/\alpha$ for individual spectra based on Equation 1. The two [\ion{O}{3}] doublet emission lines intrinsically have the identical line profiles since both transitions are from the same excited energy level of double-ionized oxygen. In reality, they could have different profiles in an observed spectrum due to different spectral resolutions and dispersions in different wavelength ranges. In particular, the resolution of the NIRSpec gratings changes rapidly with wavelength (Figure \ref{fig:resolution}). We cannot fix the flux ratio of the two lines either, because this ratio varies in different galaxies (see more details regarding the flux ratio later). Therefore, we treat the two [\ion{O}{3}] lines separately. We use a single Gaussian profile to fit individual [\ion{O}{3}] lines and obtain their line wavelengths from the best fits (Figure \ref{fig:1gfit}). 

For each spectrum in our sample, we first shift it to the rest frame using the redshift from the object catalog provided by the data product teams or from our own calculations. We then trim the spectrum and keep a wavelength range from 4920 to 5050 \AA. This baseline is long enough for us to derive a robust continuum. If either edge of a spectrum is too close ($<30$ \AA\ in the rest frame) to either of the two [\ion{O}{3}] lines, this spectrum is excluded. With the trimmed spectrum, we model its continuum and two emission lines in two iterations. In the first iteration, we assume that the line width (full-width at half maximum; FWHM) is determined by the wavelength resolution. We will see below that the majority of the lines here are indeed not resolved or barely resolved. To model the continuum, we exclude two wavelength ranges that are within $\rm 2.5\times FWHM$ from the two line centers. We then fit a second-order polynomial curve to the remaining data points that are completely dominated by the continuum emission. Traditionally we assume that a galaxy continuum has a power-law shape. However, the majority of the galaxies in our sample (i.e., strong line emitters at high redshift) have very weak continuum emission that is invisible or barely seen in the NIRSpec spectra (two examples are shown in Figure \ref{fig:1gfit}). Their continuum data points are close to zero, with many negative values, so a power-law shape cannot be used.

After the continuum subtraction, we fit a single Gaussian to the two [\ion{O}{3}] doublet lines independently. Figure \ref{fig:1gfit} shows the examples of our best fits to two spectra. For each line in a spectrum, the wavelength range chosen to fit is $\rm \pm\ 1.5\times FWHM$ from the line center. If this range covers less than 7 pixels, we increase the range to $\rm \pm\ 2.0\times FWHM$. Typically only 3--5 pixels are significantly ($>5\sigma$) above the continuum, as seen in Figure \ref{fig:1gfit}. Because of this reason, we do not use more complex functions that have more free parameters. In the second iteration, we repeat the above procedure. The only difference is that the input FWHM values of the lines are from the first iteration.

The usage of a single Gaussian does not have a definitive physical meaning. If a gas cloud has one bulk speed, a Gaussian is a good approximation to describe its line profile. An emission line from a galaxy that we observe is the superposition of multiple gas clouds with different bulk velocities, but the individual components usually cannot be resolved in the observed spectrum of the line. Therefore, the intrinsic profile of a line is more or less asymmetric. If the distribution of individual gas clouds is random and the line is relatively narrow (typical emission lines from galaxies), the overall line shape can still be approximated as a Gaussian. In our above fitting procedure, we find that a small fraction of lines have very large reduced-$\chi^2$ ($\chi^2_{\nu}$) values (up to a few hundred) from the best fits. Our visual inspection suggests that the best fits look reliable. The reason is that these lines have very high S/Ns and their line shapes slightly deviate from Gaussian, which result in large $\chi^2_{\nu}$ values. In these cases, the measured wavelength errors are significantly underestimated. This issue would appear for all other lines if their S/Ns are high enough. One may add more components in the fitting procedure to solve or alleviate the issue. However, due to the low resolution and sparse sampling of the spectra, it is not viable to add more components for the lines here. Therefore, we downgrade the spectral quality for all lines with $\chi^2_{\nu}>1.5$ by increasing their spectral errors so that their $\chi^2_{\nu}$ values of the best fits are equal to 1.5. This step does not introduce any bias. 
 
\begin{figure}[t]   
\includegraphics[width=0.45\textwidth]{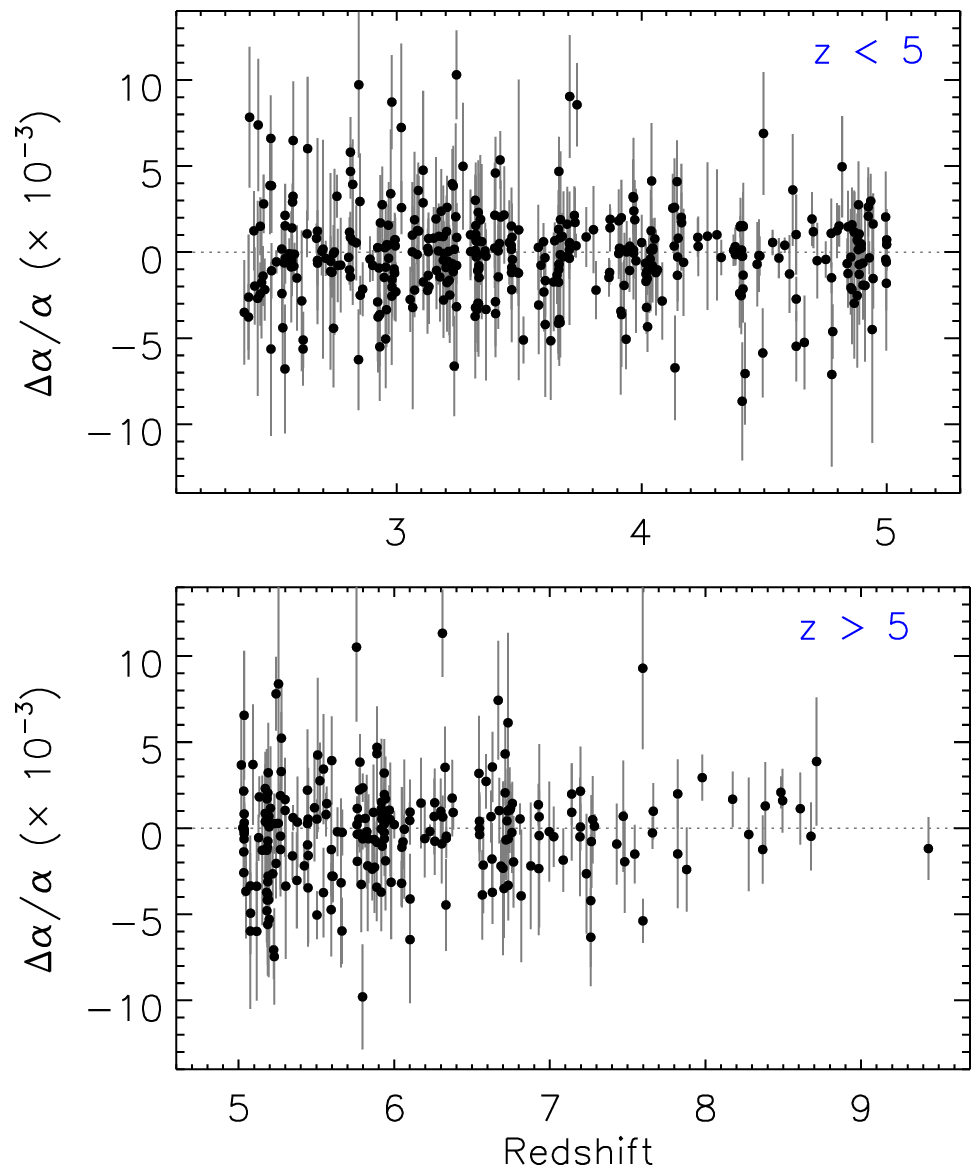}
\caption{Measured $\Delta\alpha/\alpha$ values with $1\sigma$ errors from all individual spectra. The dashed lines indicate the zero value. \label{fig:individual} }
\end{figure}

With the measured wavelengths above, we calculate $\Delta\alpha/\alpha$ using Equation 1. The measurement error is estimated by Equation 1 from the propagation of the errors of the two line wavelengths. We also take into account the systematic errors using the results derived in Section 3.2. Figure \ref{fig:individual} shows the measured $\Delta\alpha/\alpha$ values for all individual spectra. These values are also listed in the last column of Table \ref{tab:sample}. The whole table is available on line. Figure \ref{fig:alphaDistribution} shows the distributions of the $\Delta\alpha/\alpha$ values in six redshift ranges. The distributions are generally symmetric. In the two figures, some $\Delta\alpha/\alpha$ values deviate from zero at $>3\sigma$. We visually inspect their [\ion{O}{3}] lines and the best fits, and they look normal. The deviation is likely caused by the low resolution and sparse sampling of the spectra, as mentioned above. It does not introduce statistical bias.

Column 7 in Table \ref{tab:sample} shows the measured FWHMs of the [\ion{O}{3}] $\lambda$5007 lines at rest frame. If we convert these values to the observed frame, we find that they are mostly consistent with the spectral resolution, suggesting that the line width is dominated by the instrumental broadening. This agrees with the fact that the typical line width of galaxies is roughly $\rm 100-300\ km\ s^{-1}$. Column 8 in Table \ref{tab:sample} shows the measured flux ratios of the [\ion{O}{3}] doublet lines, and their relative uncertainties are mostly between 10\% and 20\%, depending on the S/Ns of the spectra. The mean and standard deviation of the ratios are 2.93 and 0.35, respectively. This flux ratio can be significantly deviate from the nominal value 3 for reasons that are not fully understood. But here the measurement uncertainties dominate the discrepancy between the observed and theoretical values.

\begin{figure}[t]   
\includegraphics[width=0.45\textwidth]{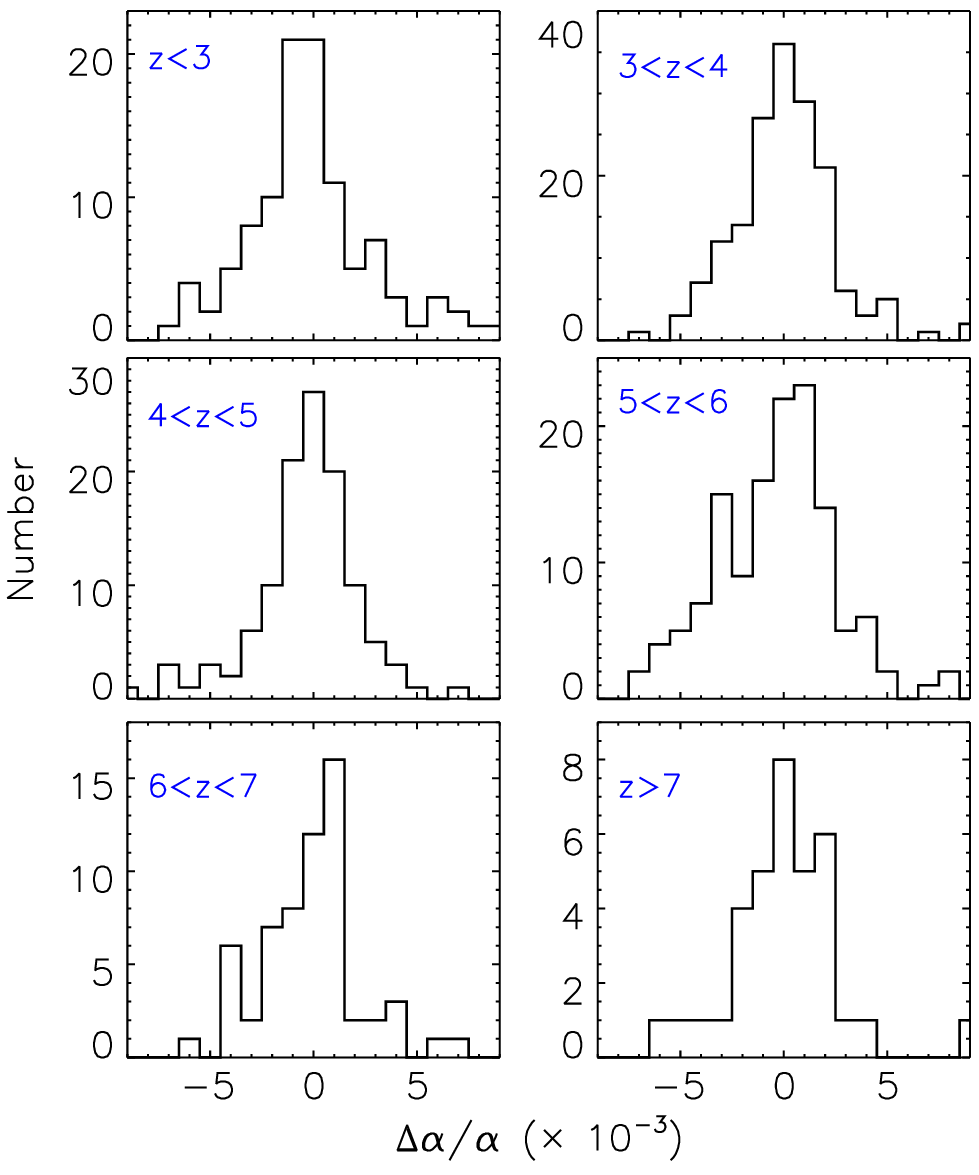}
\caption{Distributions of $\Delta\alpha/\alpha$ for individual spectra in six redshift ranges. The distributions are generally symmetric. \label{fig:alphaDistribution} }
\end{figure}

As mentioned in Section 2.3, 42 galaxies have two spectra and one galaxy has three spectra. We use these spectra to check the consistency of our $\Delta\alpha/\alpha$ measurement. The upper panel of Figure \ref{fig:multiObs} displays the differences of the measured $\Delta\alpha/\alpha$ values between the two sets of the spectra. The distribution of the differences is shown as the blue histogram in the lower panel of Figure \ref{fig:multiObs}. As we do for Figure \ref{fig:wCalib2}, we perform Monte Carlo simulations to check whether the errors in the upper panel account for the scatter of the distribution. The gray histograms illustrate 100 simulated results, and they agree well  with the observed distribution in blue. Therefore, the results from the multiple observations are well consistent.

\begin{figure}[t]   
\includegraphics[width=0.45\textwidth]{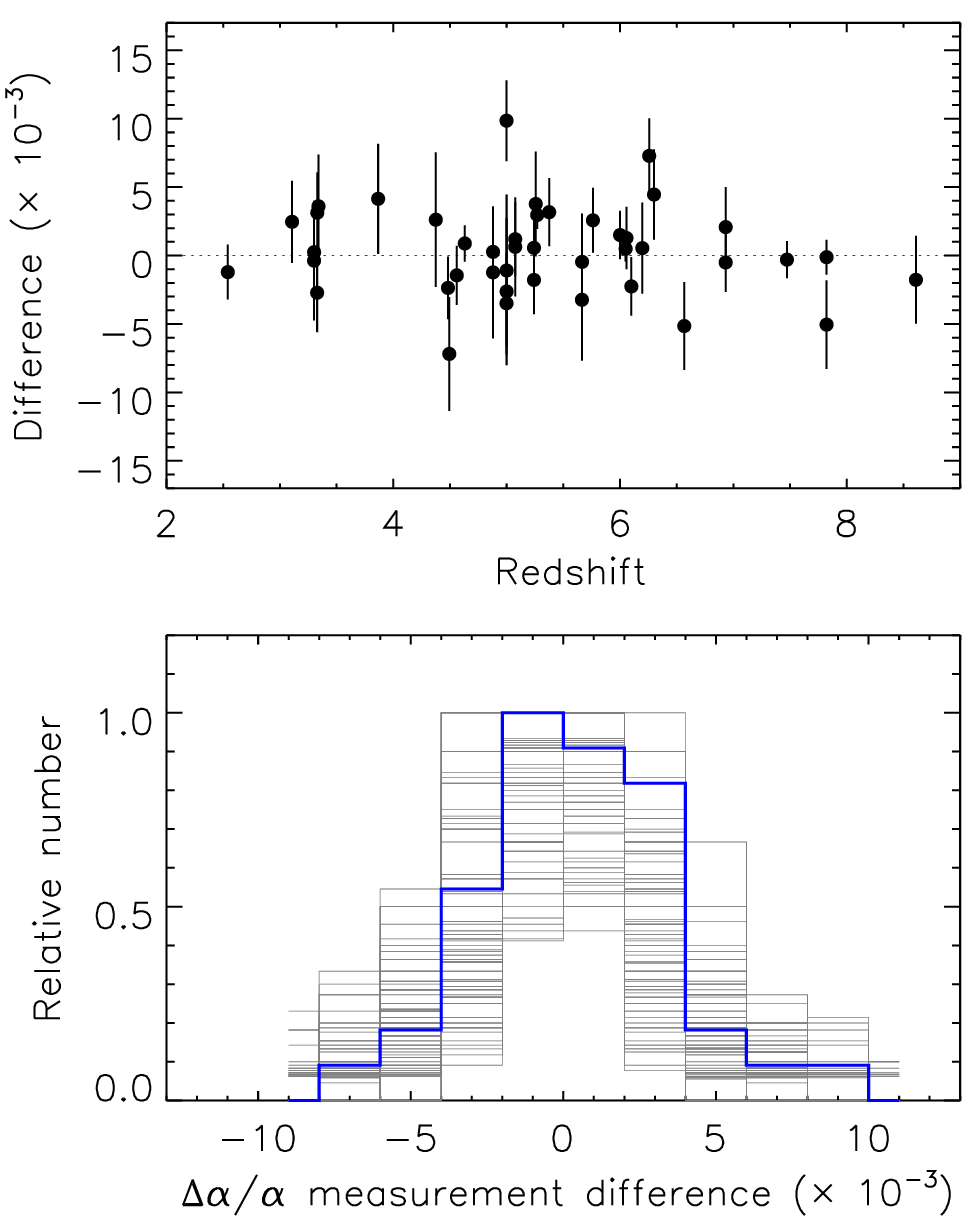}
\caption{Comparison of the measured $\Delta\alpha/\alpha$ values from the multiple observations of the same galaxy sample. The upper panel shows the differences of the  $\Delta\alpha/\alpha$ values between the two sets of the spectra. In the lower panel, the blue histogram shows the distribution of the differences from the upper panel. The gray histograms represent 100 simulated results and they are well consistent with the observed result.
\label{fig:multiObs} }
\end{figure}

\subsection{Time Variation of $\alpha$} \label{subsec:time_vary}

\begin{figure}[t]   
\includegraphics[width=0.45\textwidth]{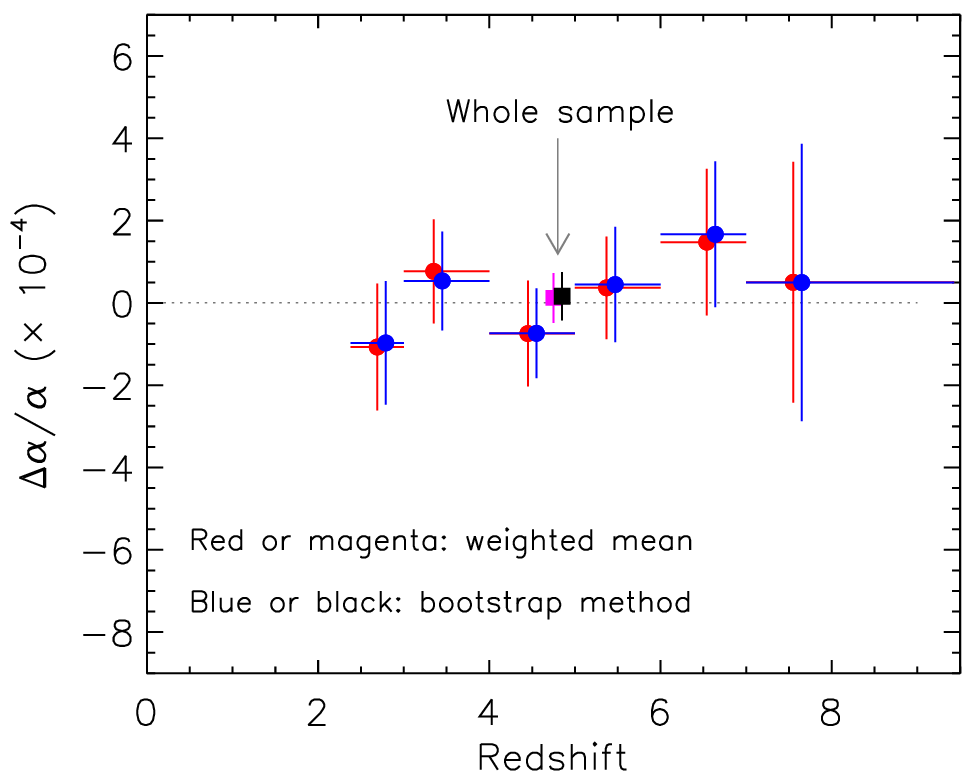}
\caption{Relative $\alpha$ variation $\Delta\alpha/\alpha$ with redshift. The red and blue circles represent the results from the weighted average and bootstrap estimate methods, respectively. They are placed at the median redshifts of the individual bins. The vertical bars indicate $1\sigma$ errors, and the horizontal bars indicate the redshift ranges covered by the individual bins. The results from the two methods are well consistent. The magenta and black squares, placed at $z=4.8$, represent the result for the whole sample. The figure shows no apparent $\alpha$ variation (within $1\sigma$) with time. \label{fig:timeVary} }
\end{figure}

In order to estimate the time variation of $\alpha$, we divide the sample into six subsamples in the following six redshift bins, $z<3$, $3<z<4$, $4<z<5$, $5<z<6$, $6<z<7$, and $z>7$. For each redshift bin, we calculate $\Delta\alpha/\alpha$ and its error using two methods. In the first method, we take a weighted mean as defined in Section 3.2. In the second method, we use a bootstrap estimate \citep[e.g.,][]{Bahcall2004}. For a redshift bin with $n$ galaxies, its $\Delta\alpha/\alpha$ and error are estimated from 10,000 simulated samples. To generate a simulated sample, we randomly draw $n$ galaxies with replacement from the real galaxies in this redshift bin, and calculate the weighted average of $\Delta\alpha/\alpha$ for the simulated sample. Then the $\Delta\alpha/\alpha$ and error values of this redshift bin are the mean and standard deviation of the 10,000 simulated values. 

Figure \ref{fig:timeVary} displays the results from the two methods. The data points are placed at the median redshifts of the individual bins. We also calculate $\Delta\alpha/\alpha$ for the whole sample and the values are $(0.12\pm0.61)\times10^{-4}$ and $(0.16\pm0.73)\times10^{-4}$ from the two methods, respectively. The magenta and black squares indicate the two measurements. They are placed at $z=4.8$ in Figure \ref{fig:timeVary}. We do not display them at the median redshift of 4.3 for a better visibility in the figure. The figure shows that the results from the two methods agree well with each other.

Figure \ref{fig:timeVary} shows that the $1\sigma$ errors of $\Delta\alpha/\alpha$ in individual redshift bins are roughly $(1.3-2.2)\times10^{-4}$, except those in the highest-redshift bin that are roughly $(3.3-4.2)\times10^{-4}$, depending on the methods adopted. It further suggests a non-variation of $\alpha$ with redshift within $1\sigma$. With the whole sample, the constraint is improved to be $\Delta\alpha/\alpha=(0.16\pm0.73)\times10^{-4}$ from the bootstrap method. These constraints are not as strong as those from observations of lower-redshift objects, particularly quasar absorption lines. However, previous studies rarely reached the redshift range of $z>4$, and here we provide the first constraints on the $\alpha$ variation at the highest redshifts up to $z>9$. Future JWST observations will provide more stringent constraint.

\subsection{Spatial Variation of $\alpha$}

We briefly test the spatial variation of $\alpha$. Some previous studies found tentative evidence for a possible variation of $\Delta\alpha/\alpha$ in space at a level below $10^{-5}$ \citep[e.g.,][]{Webb2011,King2012}. In this work, we do not have a sufficiently large sample or a sufficiently high accuracy of $\Delta\alpha/\alpha$ to perform a detailed analysis. Instead, we compare four $\Delta\alpha/\alpha$ values for four fields, including CEERS (14h20m, +52d58m), GOODS-S (03h32m, $-$27d49m), GOODS-N (12h36m, +62d15m), and A2744  (00h14m, $-$30d25m). The spectra in these fields dominate our sample. The fields point to very different directions of the sky. We calculate their $\Delta\alpha/\alpha$ values using the bootstrap method and the result is shown in Figure \ref{fig:spaceVary}. The calculated $\Delta\alpha/\alpha$ values are statistically consistent with zero, suggesting no spatial variation of $\alpha$ at a $1\sigma$ level of $\sim 2\times10^{-4}$ or above.

\begin{figure}[t]   
\includegraphics[width=0.45\textwidth]{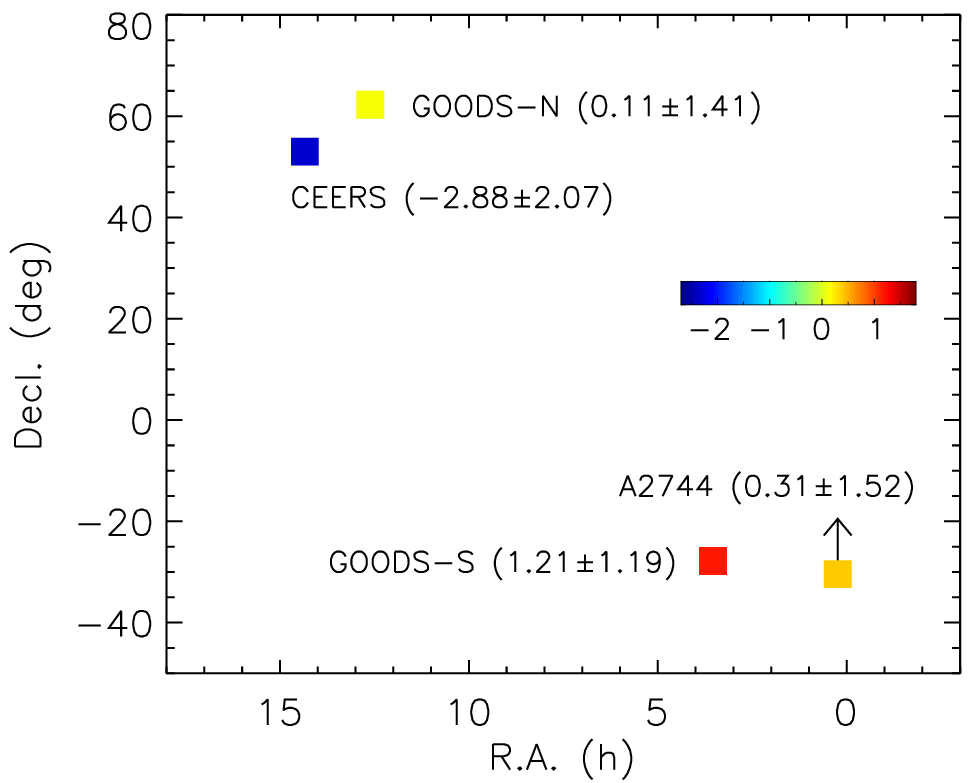}
\caption{Test of spatial variation of $\Delta\alpha/\alpha$. The color-coded squares represent the results from the bootstrap method for four fields,  including CEERS, GOODS-S, GOODS-N, and A2744. The measured $\Delta\alpha/\alpha$ values are next to the field names, with the units of $10^{-4}$. The figure shows that the four $\Delta\alpha/\alpha$ values are statistically consistent with zero within $1\sigma$. \label{fig:spaceVary} }
\end{figure}

\section{Discussion and summary} \label{sec:summary}

We have used a sample of high-redshift [\ion{O}{3}] emission-line galaxies to constrain the spacetime variation of the fine-structure constant $\alpha$ at redshift $2.5 \le z<9.5$. The sample consists of 621 JWST NIRSpec spectra with high S/Ns from 578 galaxies, including 232 spectra at $z>5$. JWST for the first time allows us to probe the $\alpha$ variation using the highest-redshift astrophysical objects. We paid a close attention to the wavelength measurement of emission lines, and performed a comprehensive analysis of the wavelength calibration for the NIRSpec spectra. We found that the calibration is generally robust and accurate for this work.

To study the time variation, the galaxy sample was divided into six subsamples based on redshift, and $\Delta\alpha/\alpha$ was calculated for the individual subsamples using two methods. The results from the two methods agree with each other. We found that the calculated $\Delta\alpha/\alpha$ values are all consistent with zero within $1\sigma$ error of $(1-2) \times10^{-4}$, depending on redshifts or subsamples. A stronger constraint was achieved to be $\Delta\alpha/\alpha = (0.2\pm0.7) \times10^{-4}$ when the whole sample was used. Our results suggest no strong variation in $\alpha$ over a wide redshift range up to $z\sim9.5$. This constraint is not as strong as those from previous studies, but this is the first constraint on $\alpha$ using the highest-redshift objects. We also briefly tested the spatial variation in $\alpha$ using galaxies in four directions on the sky and the $\Delta\alpha/\alpha$ values in different directions are also consistent with zero at a $1\sigma$ level of $\sim 2\times 10^{-4}$. This is consistent with the recent work by \citet{Jiang2024}, who used a large number of galaxies over a few thousand deg$^2$ at $z<1$ and did not find an apparent variation of $\alpha$ at a level of $10^{-4}$. 

The data in this work are mostly from JWST Cycle 1 programs, and the current constraint from the whole sample is 70 ppm ($1\sigma$; null result). Assuming that JWST will achieve 20 times more data within an expected lifetime of 20 years, statistically, the constraint on $\Delta\alpha/\alpha$ would improve to be around 15 ppm. However, this is unlikely because the systematics of the wavelength calibration will start to dominate the wavelength measurement, as we extensively discussed in Section \ref{sec:waveCal}. On the other hand, there is still room to improve the wavelength calibration by observing more calibration objects (particularly bright objects with many strong emission lines). As JWST is accumulating data at a fast pace,  we expect to have a larger dataset and put stronger constraints on $\Delta\alpha/\alpha$ in the near future.

\begin{acknowledgments}

We acknowledge support from the National Science Foundation of China (12225301) and the National Key R\&D Program of China (2022YFF0503401). SEIB is supported by the Deutsche Forschungsgemeinschaft (DFG) under Emmy Noether grant number BO 5771/1-1. Some of the data products presented herein were retrieved from the Dawn JWST Archive (DJA). DJA is an initiative of the Cosmic Dawn Center (DAWN), which is funded by the Danish National Research Foundation under grant DNRF140.

The JWST data of the JADES and CEERS programs presented in this article were obtained from the Mikulski Archive for Space Telescopes (MAST) at the Space Telescope Science Institute. The specific observations analyzed can be accessed via \dataset[doi:10.17909/8tdj-8n28]{https://doi.org/10.17909/8tdj-8n28} and \dataset[doi:10.17909/z7p0-8481]{https://doi.org/10.17909/z7p0-8481}. The DJA spectra were retrieved from {\url{https://dawn-cph.github.io/dja/}}.

\end{acknowledgments}

\vspace{5mm}
\facilities{JWST (NIRCam and NIRSPEC)}


\bibliography{ms_alpha}{}
\bibliographystyle{aasjournal}


\end{document}